\documentclass[10pt,twocolumn,aps,pra,showpacs,superscriptaddress,floatfix,nofootinbib,longbibliography]{revtex4-1}

\usepackage[T1]{fontenc}
\usepackage{graphicx}

\usepackage{amssymb}
\usepackage{amsmath}
\usepackage{color}
\usepackage{psfrag}
\usepackage{epsfig}
\usepackage{bbm}
\usepackage{bm}
\usepackage{hyperref}
\usepackage[normalem]{ulem}
\usepackage{amsthm}
\usepackage{epstopdf}

\newcommand{\ket}[1]{| #1 \rangle}
\newcommand{\bra}[1]{\langle #1 |}

\newcommand{\beq}{\begin{eqnarray}}
\newcommand{\eeq}{\end{eqnarray}}

\newcommand{\rab}{{\rho_{ AB}}}

\definecolor{nred}{rgb}{0.9,0.1,0.1}
\definecolor{nblack}{rgb}{0,0,0}
\definecolor{nblue}{rgb}{0.2,0.2,0.8}
\definecolor{ngreen}{rgb}{0.2,0.6,0.2}

\newcommand{\red}[1]{{\color{nblack} #1}}
\newcommand{\green}[1]{{\color{nblack} #1}}

\newcommand{\SMDI}{{\mathcal{S}_1}}
\newcommand{\SMDILB}{{\mathcal{S}_1^{\text{\tiny LB}}(\mathbf{P})}}
\newcommand{\QMDI}{{\mathcal{S}^{\text{\tiny LB}}(\mathbf{P})}}
\newcommand{\SbMDILB}{{\mathcal{S}_b^{\text{\tiny LB}}(\mathbf{P})}}

\newcommand{\SOMDI}{{\mathcal{W}_1}}
\newcommand{\SOMDILB}{{\mathcal{W}_1^{\text{\tiny LB}}(\mathbf{P})}}

\newcommand{\QMDIxi}{{\mathcal{S}_\xi^{\text{\tiny LB}}(\mathbf{P},\{\xi_b\})}}

\newcommand{\EBB}{{E_1}}
\newcommand{\PhiBB}{{\Phi^{BB_{0}}_+}}

\DeclareMathOperator{\tr}{Tr}

\theoremstyle{definition}

\begin{document}

\title{Experimental demonstration of measurement-device-independent measure\\ of quantum steering}

\author{Yuan-Yuan Zhao}
\thanks{These two authors contributed equally to this work}
\affiliation{Key Laboratory of Quantum Information, University of Science and Technology of China, CAS, Hefei, 230026, China}
\affiliation{Center for Quantum Computing, Peng Cheng Laboratory, Shenzhen 518055, China}
\affiliation{CAS Center for Excellence in Quantum Information and Quantum Physics,
University of Science and Technology of China, Hefei, 230026, P. R. China}

\author{Huan-Yu Ku\normalfont\textsuperscript{*,}}
\email{huan_yu@phys.ncku.edu.tw}
\affiliation{Theoretical Quantum Physics Laboratory, RIKEN Cluster for Pioneering Research, Wako-shi, Saitama 351-0198, Japan}
\affiliation{Department of Physics and Center for Quantum Frontiers of Research \& Technology (QFort), National Cheng Kung University, Tainan 701, Taiwan}

\author{Shin-Liang Chen}
\email{shin.liang.chen@phys.ncku.edu.tw}
\affiliation{Department of Physics and Center for Quantum Frontiers of Research \& Technology (QFort), National Cheng Kung University, Tainan 701, Taiwan}
\affiliation{Dahlem Center for Complex Quantum Systems, Freie Universit\"at Berlin, 14195 Berlin, Germany}

\author{Hong-Bin Chen}
\affiliation{Department of Engineering Science and Center for Quantum Frontiers of Research \& Technology (QFort), National Cheng Kung University, Tainan 70101, Taiwan}

\author{Franco Nori}
\affiliation{Theoretical Quantum Physics Laboratory, RIKEN Cluster for Pioneering Research, Wako-shi, Saitama 351-0198, Japan}
\affiliation{Department of Physics, The University of Michigan, Ann Arbor, 48109-1040 Michigan, USA}

\author{Guo-Yong Xiang}
\email{gyxiang@ustc.edu.cn}
\affiliation{Key Laboratory of Quantum Information, University of Science and Technology of China, CAS, Hefei, 230026, China}
\affiliation{CAS Center for Excellence in Quantum Information and Quantum Physics,
University of Science and Technology of China, Hefei, 230026, P. R. China}

\author{Chuan-Feng Li}
\affiliation{Key Laboratory of Quantum Information, University of Science and Technology of China, CAS, Hefei, 230026, China}
\affiliation{CAS Center for Excellence in Quantum Information and Quantum Physics,
University of Science and Technology of China, Hefei, 230026, P. R. China}

\author{Guang-Can Guo}
\affiliation{Key Laboratory of Quantum Information, University of Science and Technology of China, CAS, Hefei, 230026, China}
\affiliation{CAS Center for Excellence in Quantum Information and Quantum Physics,
University of Science and Technology of China, Hefei, 230026, P. R. China}

\author{Yueh-Nan Chen}
\email{yuehnan@mail.ncku.edu.tw}
\affiliation{Department of Physics and Center for Quantum Frontiers of Research \& Technology (QFort), National Cheng Kung University, Tainan 701, Taiwan}

\date{\today}

\begin{abstract}
Within the framework of quantum refereed steering games, quantum steerability can be certified without any assumption on the underlying state nor the measurements involved. Such a scheme is termed the measurement-device-independent (MDI) scenario. \green{Here} we introduce a measure of steerability in an MDI scenario, i.e., the result merely depends on the observed statistics and the quantum inputs. We prove that such a measure \red{satisfies the convex steering monotone.} \red{Moreover, it is robust against not only measurement biases but also losses.}
We also experimentally estimate the amount of the measure with an entangled photon source.
\red{As two by-products, our experimental results provide lower bounds on an entanglement measure of the underlying state and an incompatible measure of the involved measurement.} \red{Our research paves a way for exploring one-side device-independent quantum information processing within an MDI framework.}
\end{abstract}


\maketitle
\section{Introduction}
Entanglement~\cite{Einstein35}, steerability~\cite{Schrodinger35}, and Bell nonlocality~\cite{Bell64} are three types of quantum correlations which play essential roles in quantum cryptography, quantum teleportation, and quantum information processing~\cite{Horodecki09RMP,Brunner14RMP,UolaRev2020}. The fact that quantum steering is treated as an intermediate quantum correlation between entanglement and nonlocality leads to a hierarchical relation among them. That is, all nonlocal states are steerable, and all steerable states are entangled, but not vice versa~\cite{Wiseman07,Jones07,Quintino15} 
During the past decade, there have been many significant experimental works~\cite{Saunders10,  Bennet12,Handchen12, Smith12,Schneeloch13,Su13,Sun16, Cavalcanti09} and various theoretical results on quantum steering~\cite{Reid89,Pusey13,Walborn11,Kogias15,Costa16,Chiu2016}, including the correspondence with measurement incompatibility~\cite{Cavalcanti16,Uola14,Quintino14,Shin-Liang16c,Uola15}, one-way steering~\cite{Wollmann16,Bowles14}, temporal steering~\cite{Yueh-Nan14,Shin-Liang16,Ku16,Che-Ming15,Ku18b}, continuous-variable steering~\cite{Tatham12,Qiongyi15,Xiang17}, and measures of steering~\cite{Piani15,Skrzypczyk14,Hsieh16,Gallego15,SDPreview17,Ku18a}.

\begin{figure}[htbp]
\includegraphics[width=0.8\linewidth]{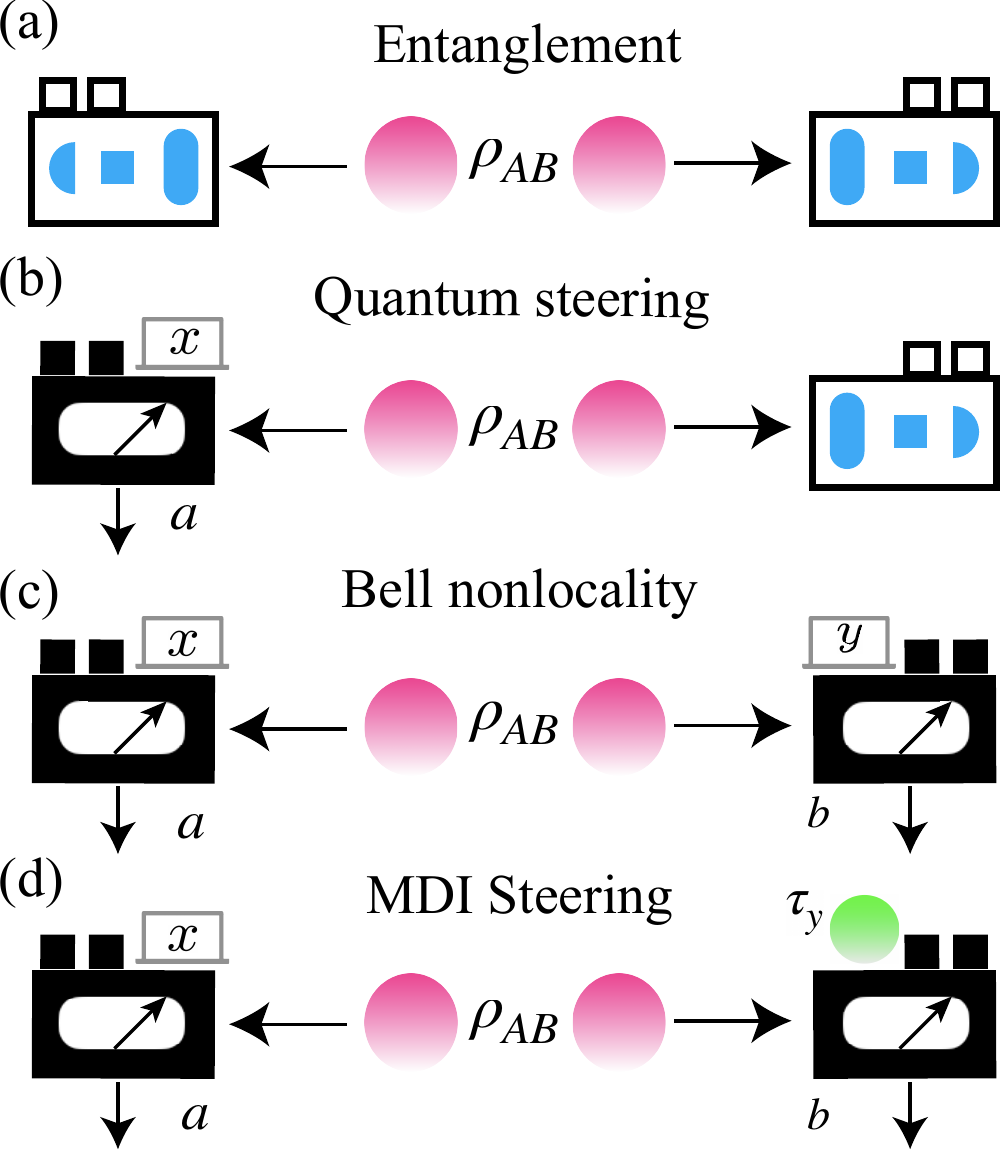}
\caption{
\red{\textbf{Schematic illustration of the entanglement, quantum steering, Bell nonlocality, and MDI steering scenarios.} A pair of entangled photons $\rab$ (pink balls) are shared between two spatially separated parties: Alice and Bob. They verify whether they share the entanglement, steering, and nonlocal resource by violating the entanglement witness, steering inequality, and Bell inequality, respectively. (a) In the entanglement certification task, Alice and Bob both perform characterized measurements (transparent box). (b) In the quantum steering scenario, one party performs uncharacterized measurements (black box) according to the classical input $\{x\}$, while the other party performs a set of characterized measurements. (c) In Bell nonlocality, Alice (Bob) receives the classical input $\{x\}$ ($\{y\}$) and returns the outcomes $\{a\}$ ($\{b\}$) with uncharacterised measurements. (d) In the MDI steering scenario, Bob's classical input $\{y\}$ of the steering scenario is replaced with quantum inputs $\{\tau_y\}$, removing the necessity of trustiness of the measurement device.}}
\label{Fig_Ent_Ste_Bell}
\end{figure}

Bell nonlocality enables one to perform the so-called \emph{device-independent} (DI) quantum information processing~\cite{Brunner14RMP,Gallego10,Bancal11,Cavalcanti12,Acin07}, i.e., one makes no assumption on the underlying state nor the measurements performed. From the hierarchical relation~\cite{Wiseman07}, it naturally leads to the fact that a Bell inequality can be treated as a \emph{DI entanglement witness}. Nevertheless, not all entangled states can be detected by using a Bell inequality violation~\cite{Werner89}. Recently, based on Buscemi's \emph{semi-quantum nonlocal games}~\cite{Francesco12}, Branciard \emph{et al.}~\cite{Branciard13} proposed a collection of entanglement witnesses in the so-called \emph{measurement-device-independent} (MDI) scenario. Compared with the standard DI scenario, there is one more assumption in an MDI scenario: the input of each detector has to be a set of tomographically complete quantum states instead of real numbers. Such a simple relaxation leads to that \emph{all} entangled states can be certified by the proposed MDI entanglement witnesses~\cite{Francesco12,Branciard13}. This characterization gives rise to the recent works providing frameworks for MDI measures of entanglement~\cite{Rosset18b,Shahandeh17,Verbanis16}, non-classical teleportation~\cite{Cavalcanti17}, and non-entanglement-breaking channel verification~\cite{Rosset18,Uola192,Yuan19}.

Recently, Cavalcanti~\emph{et al.}~\cite{Cavalcanti13} introduced another type of nonlocal game, dubbed as \emph{quantum refereed steering games} (QRSGs). In each of such games, one player, denoted as Alice, is questioned and answers with real numbers, while the other player, saying Bob, is questioned with (isolated) quantum states but still answers with real numbers. They showed that there always exists a QRSG with a higher winning probability when the players are correlated by a steerable state~\cite{Cavalcanti13}. Later, Kocsis \emph{et al.}~\cite{Kocsis15} experimentally proposed a QRSG and verified the steerability for the family of two-qubit Werner states in such a scenario, which is also referred to as an MDI scenario. \red{Moreover, such a QRSG scenario can be used to generate the private random number by maximal violation of the higher dimensional steering inequality under the MDI framework~\cite{Skrzypczyk2018,Guo2019}.}

\green{Here} we consider a variant of QRSGs, by which we propose the first MDI steering measure (MDI-SM) of the underlying unknown steerable resource without accessing any knowledge of the involved measurements.
We show that the MDI-SM is a standard measure of steerability, i.e., a \emph{convex steering monotone}~\cite{Gallego15}, by proving that it is equivalent to the previously proposed measures: the steering robustness~\cite{Piani15} and the steering fraction~\cite{Hsieh16}. Therefore, our proposed measure not only coincides with the degree of steerability of the underlying steerable resource, but also quantifies the degree of entanglement of the shared quantum state~\cite{Piani15} and incompatibility of the measurements involved~\cite{Cavalcanti16,Shin-Liang16c,Chen18}. Furthermore, MDI-SM can be computed via a semidefinite program. We also show the MDI-SM is robust, in the sense that it can detect steerability \green{in the presence of} detection losses \red{and biases}~\cite{Branciard13,Rosset18b,Shahandeh17,Verbanis16}.

Finally, we experimentally estimate the degree of steerability of the family of two-qubit Werner states in an MDI scenario.
We consider that Alice performs three qubit-measurements in the mutually unbiased bases (MUBs) since they can be used to demonstrate the strongest steerability to Bob when Alice has three \green{measurement} settings~\cite{Skrzypczyk14}. On the other hand, Bob performs the Bell-state measurement (BSM) on his part of the state and the quantum inputs. Based on the observed correlations, the steerability of the family of two-qubit Werner states are quantified by solving a semidefinite program. As mentioned before, the experimental data naturally \red{bounds} the degree of entanglement of the underlying state, and the amount of measurement incompatibility of Alice's measurements. Compared with the previous experimental works~\cite{Xu14,Kocsis15,Verbanis16,Wollmann2018,Wollmann2019} in the MDI scenarios, our method not only certifies the existence of entanglement and measurement incompatibility, but also \red{bounds} these quantities. \red{Moreover, our experimental result roughly relates with the probabilities of successful subchannel discrimination in the MDI scenario~\cite{Piani15,Sun2018}}.


\section{Measurement-device-independent measure of steerability} Through this work, we assume that all quantum states act on a finite dimensional Hilbert space $\mathcal{H}$. The sets of density matrices and operators acting on $\mathcal{H}$ are denoted by $\mathsf{D}(\mathcal{H})$ and $\mathsf{L}(\mathcal{H})$, respectively. We denote the index sets of a finite number of elements by $\mathcal{A}$, $\mathcal{B}$, $\mathcal{X}$, and $\mathcal{Y}$. The probability of a specific index, say $a\in\mathcal{A}$, is denoted by $p(a)$.

In the MDI steering scenario, we consider two spatially separated parties, Alice and Bob, sharing a quantum state $\rho_{AB}\in \mathsf{D}(\mathcal{H}_{A}\otimes \mathcal{H}_{B})$ (see Fig.~\ref{Fig_Ent_Ste_Bell}). During each round of the experiment, Alice receives a classical input $x\in\mathcal{X}$ and performs the corresponding measurement on her system with an outcome $a\in\mathcal{A}$. On the other hand, Bob performs a joint measurement on his system and a trusted input quantum state $\tau_y\in \mathsf{D}( \mathcal{H}_{B_0})$, with $y\in\mathcal{Y} $. 
We note that the trustiness represents the state is well prepared and there is no side channel to transmit the state information.
Their joint probability distributions can be expressed as:
$p(a,b|x,\tau_y) = \text{Tr}\left[\left(E_{a|x}\otimes E_b\right)\left(\rho_{AB}\otimes\tau_y\right)\right]\quad \forall a,b,x,y,$
where $\{E_{a|x}\}_a\in \mathsf{L}(\mathcal{H}_{A})$ and $\{E_b\}_b\in\mathsf{L}(\mathcal{H}_{B}\otimes\mathcal{H}_{B_{0}})$ are the positive-operator valued measurements (POVM) (i.e., the general quantum measurements) describing Alice's and Bob's measurement with the corresponding outcomes $\{a\}$ and $\{b\}$, respectively.

\green{Within} the framework of the resource theory of quantum steering~\cite{Gallego15}, we concern more about the underlying \emph{assemblage}~\cite{Pusey13} Bob receives rather than the shared quantum state. That is, we describe the obtained correlation by Bob's joint measurement $\{E_b\}$ on the quantum inputs $\{\tau_y\}$ and the assemblage $\{\sigma_{a|x}\}$:
\begin{equation}
p(a,b|x,\tau_y)=\text{Tr}\left[E_b (\sigma_{a|x}\otimes\tau_y)\right].
\label{pabxy}
\end{equation}
An assemblage $\{\sigma_{a|x}\}$ is a set of subnormalized quantum states defined by $\sigma_{a|x}=\text{Tr}_A(\rho_{AB}~E_{a|x}\otimes\openone)$~\cite{Pusey13}, which includes both the information of Alice's marginal statistics $p(a|x)=\text{tr}(\sigma_{a|x})$ and the normalized states $\hat{\sigma}_{a|x}=\sigma_{a|x}/p(a|x)\in \mathsf{D}(\mathcal{H}_{B})$ Bob receives. The free state of the quantum steering (denoted as unsteerable assemblage) is the assemblage admitting a local-hidden-state (LHS) model~\cite{Wiseman07}, described by a deterministic strategy $D(a|x,\lambda)$ and pre-existing (subnormalized) quantum states $\{\sigma_\lambda\}$, such that $\sigma_{a|x}=\sigma^{\text{US}}_{a|x}=\sum_{\lambda}D(a|x,\lambda)\sigma_\lambda~\forall~a,x$. In particular, the set of all unsteerable assemblages $\mathsf{LHS}$ forms a convex set; consequently, for a given steerable assemblage $\{\sigma_{a|x}^{\rm{S}}\}$, there always exists a set of positive semidefinite operators $\{F_{a|x}\succeq 0\}$, called a \emph{steering witness}, such that $\tr\sum_{a,x}F_{a|x}\sigma_{a|x}^{\rm{S}} > \alpha$, while $\tr\sum_{a,x}F_{a|x}\sigma_{a|x}^{\rm{US}} \leq \alpha \quad \forall \{\sigma_{a|x}^{\rm{US}}\}\in\mathsf{LHS}$~\cite{Cavalcanti09,Pusey13,SDPreview17,Skrzypczyk14,Piani15}, where $\alpha:= \max_{\{\sigma_{a|x}^{\rm{US}}\}\in\mathsf{LHS}} \tr \sum_{a,x} F_{a|x}\sigma_{a|x}^{\rm{US}}$ is the local bound of the steering witness.

In what follows, we will construct the MDI-SM by using the \green{aforementioned} existence of a steering witness for any steerable assemblage. We start by considering a variant of QRSGs. Indeed, Eq.~\eqref{pabxy} can be treated as correlations obtained in a variant \green{of} QRSGs with steerable \emph{assemblages} being a resource. We stress that, in the standard QRSGs, one \green{instead} treats a set of steerable \emph{states} as a resource in such a game. \green{These two resources are inequivalent} because one can obtain the same assemblage from different states and measurements.
With this, we define \green{a payoff associated exclusively to} a single Bob's outcome \green{($b=1$)} as
\begin{equation}
\mathcal{W}\left(\mathbf{P},\beta\right)=\sum_{a,x,y}\beta^{x,y}_{a,1}p(a,1|x,\tau_y),
\label{payoff2}
\end{equation}
where $\mathbf{P}:= \{p(a,1|x,\tau_y)\}$ is the experimentally observed statistics from an assemblage $\{\sigma_{a|x}\}$ based on Eq.~\eqref{pabxy} and $\beta:=\{\beta_{a,1}^{x,y}\}$ is a set of real coefficients.

\green{With the above definition}, we prove that, given any steerable assemblage, there always exists a set of real coefficients $\beta$, such that the payoff $\mathcal{W}\left(\mathbf{P},\beta\right)$ is strictly higher than those obtained from unsteerable assemblages. \green{Details of the proof are given in the Supplementary Material.} In other words, the payoff $\mathcal{W}\left(\mathbf{P},\beta\right)$ is effectively the same as the standard steering witness, in the sense that all steerable assemblages can be faithfully verified by a properly chosen $\mathcal{W}\left(\mathbf{P},\beta\right)$.
We note that the witness $\mathcal{W}\left(\mathbf{P},\beta\right)$ can be seen as a generalization of a standard Bell inequality (see Ref.~\cite{Branciard13} for a similar formulation in the entanglement scenario), and is used to generalize the result of Ref.~\cite{Kocsis15}, wherein the family of two-qubit Werner states is explicitly considered.

Now we stand in the position to introduce the MDI-SM for an unknown assemblage $\{\sigma_{a|x}\}$, denoted by
\begin{equation}
\SMDI:=\max\left\{\SOMDI-1,0\right\},
\label{Eq_SMDI}
\end{equation}
with
\begin{equation}
\SOMDI := \sup_{\beta,\mathbf{P}}\frac{\mathcal{W}(\mathbf{P},\beta)}{\mathcal{W}_{\text{LHS}}(\beta)},
\label{Eq_SOMDI}
\end{equation}
where $\mathcal{W}_{\text{LHS}}(\beta)=\sup_{\mathbf{P}\in\mathsf{LHS}}\mathcal{W}(\mathbf{P},\beta)$ is the local bound for a given $\beta$.
The physical meaning of the proposed measure is simple and the idea is very similar to that of the nonlocality fraction~\cite{DanielCavalcanti13}: if the given correlation is unsteerable (i.e., it admits \green{an} LHS model), then $\mathcal{W}(\mathbf{P},\beta)\leq\mathcal{W}_{\text{LHS}}(\beta)$, and therefore $\SMDI=0$. On the other hand, if the correlation is steerable, $\mathcal{W}(\mathbf{P},\beta)>\mathcal{W}_{\text{LHS}}(\beta)$, then $\SMDI>0$. In the \green{Supplementary Material}, we further prove that:
\begin{itemize}
\item{$\SMDI$ is a steering monotone since it is equivalent to the steering fraction and the steering robustness.}
\item{The optimal $\mathbf{P}:=\{p(a,1|x,\tau_y) = \tr[E_1(\sigma_{a|x}\otimes\tau_y)]\}$ in Eq.~\eqref{Eq_SOMDI} is obtained when Bob's measurement is the projection onto the maximally entangled state. That is, $E_1=|\PhiBB\rangle\langle\PhiBB|$, with $|\PhiBB\rangle=1/\sqrt{d_{B}}\sum_{i=1}^{d_{B}}|i\rangle\otimes|i\rangle$.}
\end{itemize}


\green{After introducing our measure of the steerability in an MDI scenario, we proceed by considering the following two practical circumstances. First, one would like to estimate the degree of steerability of a given data table without any {\it a priori} knowledge about the experimental setup. Second, as the experimental apparatuses are inevitably erroneous in practical situations, how can one estimate the degree of steerability in the absence of the optimization of Bob's measurement? These two circumstances give rise to the attempt to estimate the degree of steerability of an experimentally observed correlation $\mathbf{P}$ when lacking the knowledge about the underlying assemblage.}

\green{In the case of an inaccessible assemblage, the optimization over $\mathbf{P}$ in Eq.~\eqref{Eq_SOMDI} becomes not feasible. Consequently, the alternative quantity}
 $\SOMDILB:=\sup_{\beta}\frac{\mathcal{W}(\mathbf{P},\beta)}{\mathcal{W}_{\text{LHS}}(\beta)}$ is a lower bound on $\SOMDI$, and
\begin{equation}
\SMDILB:=\max\Big\{\SOMDILB-1,0\Big\}
\label{Eq_QMDISW}
\end{equation}
provides a lower bound on $\SMDI$. Trivially, the bound becomes tight when Bob's measurement is the projection onto the maximally entangled state $\EBB=|\PhiBB\rangle\langle\PhiBB|$. Note that even if Bob's inputs do not form a complete set, Eq.~\eqref{Eq_QMDISW} still provides a valid lower bound~\cite{Rosset18b}. This can be understood from the fact that the set of tomographically complete inputs is a resource for Bob to demonstrate steerability in an MDI scenario. The lack of a completeness of quantum inputs can only decrease the degree of steerability.

\green{Furthermore, to underpin the practical viability of our measure, we stress that the maximal value of Eq.~\eqref{Eq_QMDISW} is computable via} a semidefinite program (see the \green{Supplementary Material} for details):
\begin{equation}
\begin{aligned}
\text{given}&~~\{p(a,1|x,\tau_y)\}~\text{and}~\{\tau_y\}\\
\max_{\tilde{\beta}} & \sum_{a,x,y}\tilde{\beta}^{x,y}_{a,1}p(a,1|x,\tau_y)-1\\
\text{s.t.}~&d\openone-\sum_{a,x,y}D(a|x,\lambda)\tilde{\beta}^{x,y}_{a,1}\tau_y\succeq 0 \quad\forall\lambda\\
&\sum_y \tilde{\beta}^{x,y}_{a,1}\tau_y\succeq 0\quad\forall a,x.
\end{aligned}
\label{Eq_quantitative_MDI_steering_SDP}
\end{equation}
This program can be \green{performed} for a given \green{experimentally} observed correlation $\textbf{P}$. Therefore, it \green{works well particularly} when Bob's measurement is the optimal one, i.e., the projection onto the maximally entangled state. In this case, the solution of Eq.~\eqref{Eq_quantitative_MDI_steering_SDP} gives the exact value of the MDI-SM defined in Eq.~\eqref{Eq_SMDI}.

Finally, we would like to show that the MDI-SM is robust against detection losses. To see this, we consider the average loss rate of Bob's measurement $\eta\in[0,1]$. The observed correlation in this case is $p_\eta(a,1|x,\tau_y)=\eta\cdot p(a,1|x,\tau_y)$, shrinking the MDI-SM by $\eta$, i.e., $\eta\cdot\SMDI$. As can be seen above, the shrinking quantity $\eta\cdot\SMDI$ is still able to detect steerability in an MDI scenario with arbitrary detection losses and provide a lower bound on the steerability of the underlying assemblage (see Refs.~\cite{Verbanis16,Branciard13} for similar discussions in the MDI entanglement scenario.)

\begin{figure*}[htbp]
\centering
\includegraphics[width=1\linewidth]{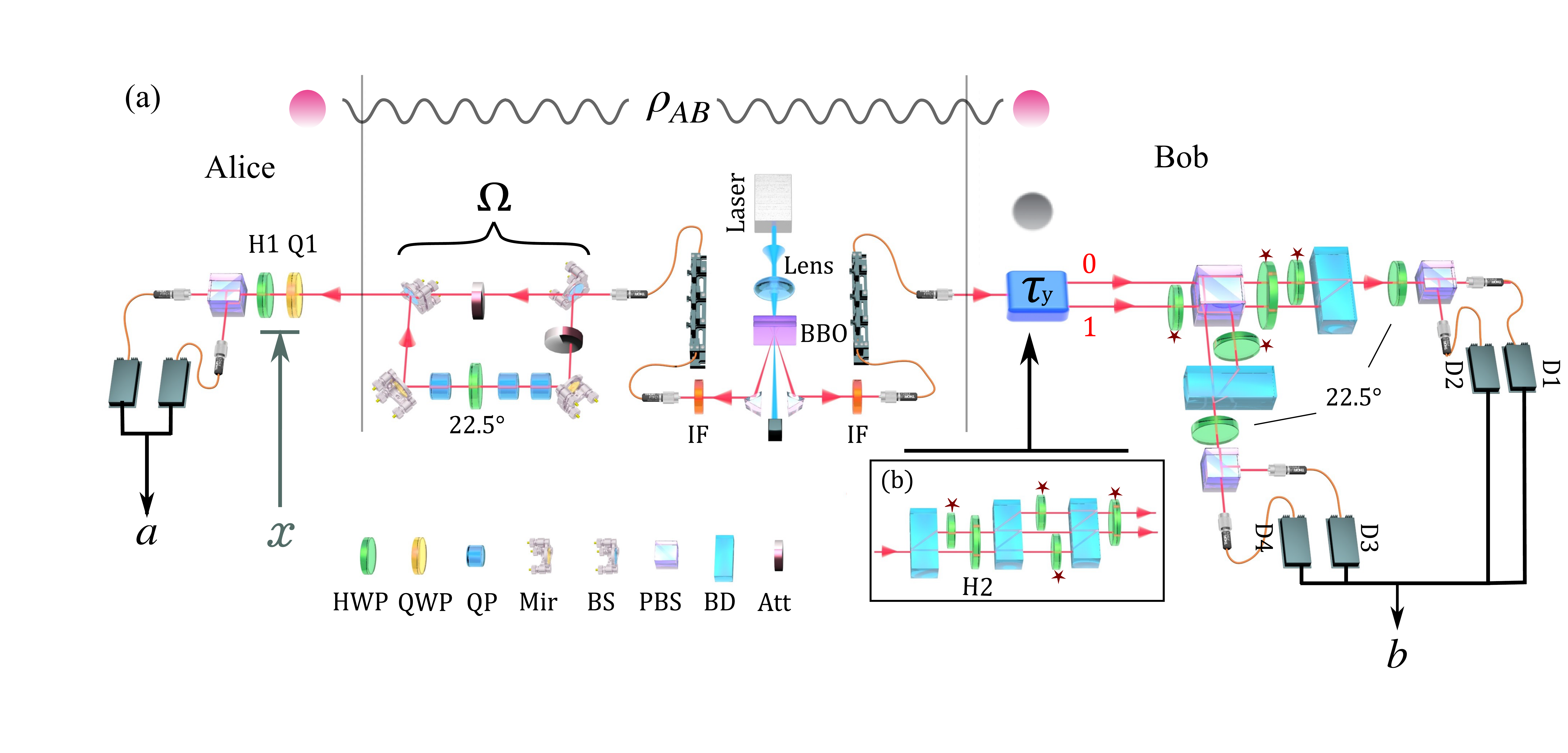}
\caption{\textbf{Schematic drawing of the experimental setup.} (a) The singlet state of a pair of photons $\ket{HV}-\ket{VH}$ is generated by a spontaneous parametric down-conversion process, where $H$ ($V$) \red{represents} the horizontally (vertically) polarized direction. The Werner state is prepared by adding white noise (denoted by $\Omega$) to the system. Then one of the photons is sent to Alice, who uses Q1, H1, and PBS to perform the measurement $x$. The other photon is sent to Bob with an additional qubit system $\tau_y$ encoded on the photon's path degree of freedom `0' and `1'. We emphasize the preparation of the trusted quantum system in \green{panel (b)}.  Now Bob performs a complete Bell-state measurement on the equivalent two-qubit systems, i.e., measuring the polarization directions and the spatial paths of the \green{single particle}, and returns an outcome $b$. At the end, a set of probability distributions $\{p(a,b|x,\tau_y)\}$ is obtained to quantify the degree of steerability of the steerable resource. Abbreviations of the components are: BBO, barium borate crystal; HWP(H), \green{half-wave} plate; IF, interference filter; Att, attenuator; Mir, mirror; QP, quartz plate; QWP(Q), \green{quarter-wave} plate; PBS, polarizing beam splitter; BS, beam splitter; BD, beam displacer. The star represents that the HWP's axis is oriented at $45^{\circ}$.}
\label{fig:setup}
\end{figure*}

\begin{figure*}[htbp]
\centering
\includegraphics[width=1\linewidth]{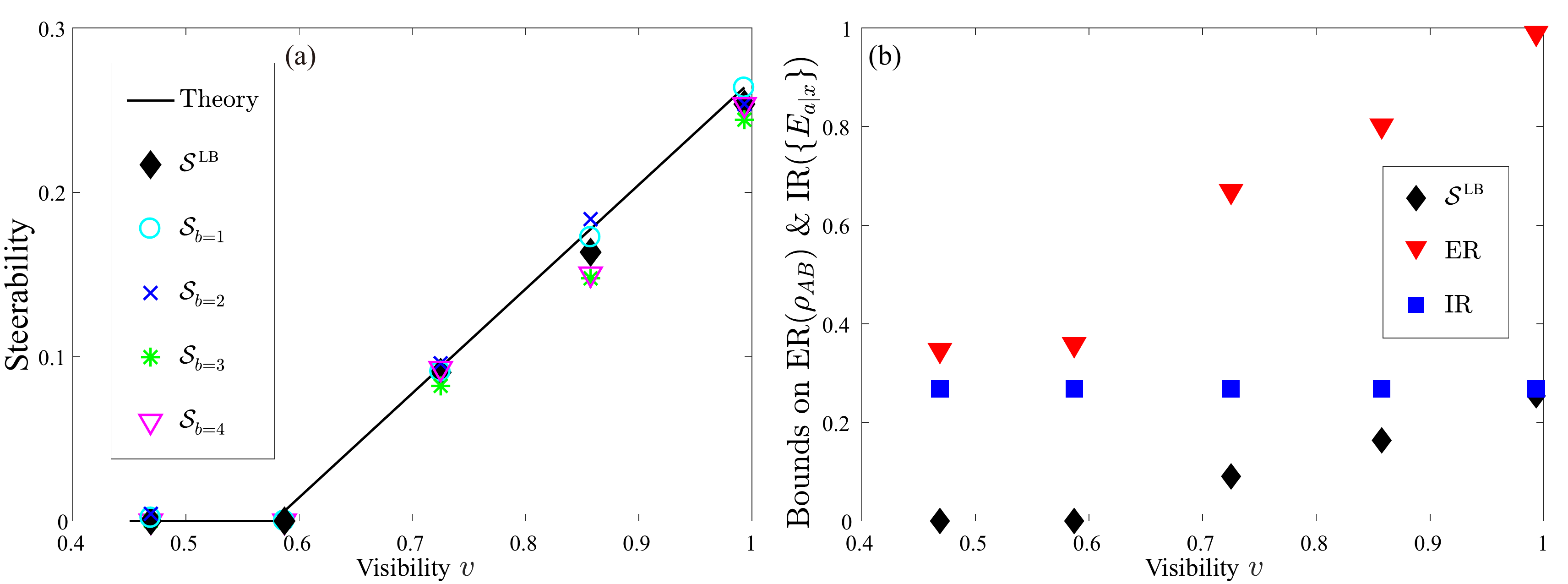}
\caption{\textbf{Results of the MDI-SM and the estimation of entanglement and measurement incompatibility.} (a) The MDI experimental demonstration of estimating steerability of the family of two-qubit Werner states when considering Alice has three measurement settings. The theoretical prediction of the MDI-SM is plotted in the black line. The tailored estimator $\QMDI$ described in Eq.~\eqref{Eq_SOMDI_qubit} for this experiment is marked as diamonds ({\color{black} $\blacklozenge$}). The MDI-SM in Eq.~\eqref{Eq_SMDI} are marked using circles ({\color[rgb]{0, 1, 1} $\bigcirc$}), crosses ({\color[rgb]{0, 0, 1} $\times$}), stars ({\color[rgb]{0, 1, 0} $*$}), and triangles ({\color[rgb]{1,0,1} $\triangledown$}).
(b) \red{MDI lower bounds on the degree of entanglement and incompatibility}. The diamond symbols ({\color{black} $\blacklozenge$}) in (a) and (b) represent the same quantity. We use the tailored estimator $\QMDI$ as lower bounds on the entanglement robustness (ER) of the underlying state and the incompatibility robustness (IR) of Alice's measurements. The actual values of these two quantities are represented by triangles ({\color{red} $\blacktriangledown$}) and squares ({\color{blue} $\blacksquare$}), respectively. By using the Monte Carlo algorithm, we obtain the standard deviations of $\SbMDILB$ in the value around $0.007$ and the standard deviations of $\QMDI$ in the value around  $0.004$ for three measurement settings by error propagation.}
\label{fig:MDI_steering_ER_IR}
\end{figure*}

\section{Experimental estimation of MDI-SM}
In the following, we will experimentally demonstrate how to estimate, in an MDI manner, the degree of steerability of the underlying steerable resource given by Alice's three measurement settings with the two dimensional MUBs acting on the two-qubit Werner states, namely $\rho_{AB}=v\ket{\psi^-}\bra{\psi^-}+\left(\frac{1-v}{4}\right)\openone$, with visibility $0\leq v\leq 1$, singlet state $\ket{\psi^-}=\frac{1}{\sqrt{2}}(\ket{HV}-\ket{VH})$, and $\openone$ being the identity operator.


The schematic diagram of our experimental setup is given in Fig.~\ref{fig:setup}. The system state is encoded on the polarization ($H$, $V$) where $H~(V)$ represents the horizontally (vertically) polarized direction of the photon. Through a spontaneous parametric down-conversion process, we generate pairs of maximally entangled photons' state $\frac{1}{\sqrt{2}}(\ket{HV}-\ket{VH})$. The Werner state is prepared by dephasing the photons to a completely mixed state with probability $(1-v)$~\cite{nonlocal2012,Qi17}. On Bob's side, a trusted device \red{shown in Fig.~\ref{fig:setup}(b)} prepares the auxiliary qubit $\tau_y$ on the path degree of freedom of his owned photon. Note that, although we encode Bob's shared state (that with Allice) and his quantum input in the same photon, these two states are indeed in different degrees of freedom. More specifically, these two states are prepared by different preparation devices, one for creating the bipartite quantum state $\rab$ while the other for generating $\tau_y$. That is to say, in our MDI scenario under consideration, the former preparation device is not trusted while the latter is trusted.

On Alice's side, she uses the \green{quarter-wave} plate Q1, the \green{half-wave} plate H1 combined with a polarization beam splitter to perform a measurement according to the value of $x$, and returns the outcome $a$ to the referee. While Bob needs to implement the optimal joint measurement, i.e., BSM on two degrees of freedom of the same particle (the polarization and the path degree of freedom), similar to the former works~\cite{bell1,bell2,Verbanis16}. This method avoids the entangled measurement on two particles, which is a tough task with 50\% efficiency in linear optics~\cite{bell4,bell5}. All the experimental details can be found in the \green{Supplementary Material}.
Moreover, a joint-measurement apparatus does not receive any information of the input quantum state before performing the measurement. More specifically, there is no side channel which transmits any information of the state to the measurement apparatus. Such protocol is physically and realistically more reliable than a situation where a referee prepares a trust quantum input to Bob. See the \green{Supplementary Material} for more experimental details.

\section{Experimental results}
To measure the steerability, we use the above experimental setup. More specifically, after sending the two-qubit Werner state $\rho_{AB}$ to Alice and Bob, we obtain the set of probability distributions $\{p(a,b|x,\tau_y)\}$ [described in Eq.~\eqref{pabxy}] by which Alice performs measurements in the Pauli bases \green{$X, Y,$ and $Z$,} on her part of the system, while Bob performs the joint measurement on his part of the system and his quantum inputs $\tau_y$. Bob's tomographically complete set of quantum inputs is composed of eigenstates of the three Pauli matrices. The joint measurement performed by Bob is the BSM, i.e., the optimal measurement, so that the value of the measure $\SMDI$ can be achieved.

Due to our experimental setup, we further show that for the underlying assemblage $\{\sigma_{a|x}\}$ being a qubit, all of the four measurement operators $\{E_b\}_{b=1,2,3,4}$ of the BSM are optimal for Bob, i.e., the produced correlation for each $b$ leads to the maximum value of Eq.~\eqref{Eq_SOMDI}. \red{Details of the discussion of the two-qubit case is given in the Supplementary Material.} Therefore, Eq.~\eqref{Eq_QMDISW} can be modified into the following form:
\begin{equation}
\QMDI := \max\left\{\frac{1}{4}\sum_{b=1}^4 {\mathcal{W}}_b^{\text{\tiny LB}}(\mathbf{P})-1,0\right\},
\label{Eq_SOMDI_qubit}
\end{equation}
where ${\mathcal{W}}_b^{\text{\tiny LB}}(\mathbf{P}):=\sup_{\beta'}\frac{\mathcal{W}(\mathbf{P},\beta')}{\mathcal{W}_{\text{LHS}}(\beta')}$ with ${\beta'}:=\{\beta_{a,b}^{x,y}\}_{a,x,y}$ for each $b$.
When there is a detection bias between the four detectors of the BSM, Eq.~\eqref{Eq_SOMDI_qubit} also provide a valid lower bound on the proposed measure. More specifically, consider that we have four detectors with the biased detection rates of $\xi_1$, $\xi_2$, $\xi_3$, and $\xi_4$, respectively, with $\sum_b \xi_b=4$ and $\xi_b\geq 0$~$\forall b$. For the ideal case, $\xi_b=1$ for all $b$. When there exists some bias, the observed correlation will be $\xi_b\cdot p(a,b|x,\tau_y)$.
Obviously, this correlation also reveals the steerability of the underlying resource, i.e.,
\begin{equation}
\begin{aligned}
&\QMDIxi:=\max\left\{\frac{1}{4}\sum_{b=1}^4 \mathcal{W}_{b,\xi}^{\text{\tiny LB}}(\mathbf{P},\{\xi_b\})-1,0\right\}\\
&:=\max\left\{\frac{1}{4}\sum_{b=1}^4 \xi_b\cdot\sum_{axy}\beta_{a,b}^{*,x,y}p(a,b|x,\tau_y)-\xi_b,0\right\}\\
&\leq\max\left\{\frac{1}{4}\sum_{b=1}^4\xi_b \mathcal{W}_b^{\text{\tiny LB}}(\mathbf{P})-1,0\right\}\\
&=\QMDI,
\end{aligned}
\label{Eq_over}
\end{equation}
where $\beta_{a,b}^{*,x,y}$ is the optimal set of coefficients for the biased correlation $\xi_b\cdot p(a,b|x,\tau_y)$.

Our experimental estimation on $\SMDI$ is plotted in Fig.~\ref{fig:MDI_steering_ER_IR} (a). As can be seen, although $\QMDI$ in Eq.~\eqref{Eq_SOMDI_qubit} may not perform the best among the other fine-grained terms $\SbMDILB:={\mathcal{W}}_b^{\text{\tiny LB}}(\mathbf{P})-1$, it is the most suitable one in the sense that the variance from the theoretical prediction is the smallest. Besides, some fine-grained terms wrongly detect the existence of steerability due to the overestimation caused by the detection bias (i.e., the estimation of steerability in Fig.~\ref{fig:MDI_steering_ER_IR}(a) when the visibility is lower than $1/\sqrt{3}$). With Eq.~\eqref{Eq_over}, such overestimation will not occur when we use the quantity $\mathcal{S}^{\text{\tiny LB}}(\mathbb{P})$. Therefore, our estimation on the MDI-SM is robust against not only detection biases but also losses.

Except for estimating the degree of steerability of the underlying assemblage in an MDI scenario, here we show that our experimental results directly \red{bound} the degree of entanglement $\text{ER}(\rab)$ of the underlying state and the degree of measurement incompatibility $\text{IR}(\{E_{a|x}\})$ of Alice's measurements. We \red{briefly} recall these two quantities in the \red{Supplementary Material}. The result is shown in Fig.~\ref{fig:MDI_steering_ER_IR} (b). The detail of the quantum state tomography to access these two quantities are also shown in the \red{Supplementary Material}. Our results are based on the fact that the steering robustness of the assemblage $\text{SR}(\{\sigma_{a|x}\})$ is a lower bound on the entanglement robustness $\text{ER}(\rab)$~\cite{Piani15} and incompatibility robustness $\text{IR}(\{E_{a|x}\})$~\cite{Cavalcanti16,Shin-Liang16c,Chen18}. Therefore, as $\QMDI$ is a lower bound on the steering robustness, $\QMDI$ is also used to provide a lower bound on $\text{ER}(\rab)$ and $\text{IR}(\{E_{a|x}\})$.

\section{Concluding Remarks}
In this work, we consider a variant of quantum refereed steering games (QRSGs), by which we introduce a measure of steerability in a measurement-device-independent (MDI) scenario, i.e., without making assumptions on the involved measurements nor the underlying assemblage. The only characterized quantities are the observed statistics and a tomographically complete set of quantum states for Bob. Through this, all steerable assemblages can be witnessed, in contrast to the fact that only a subset of steerable assemblages can be detected in the standard device-independent (DI) scenario.
We further show that it is a convex steering monotone by proving the equivalence to the steering fraction as well as the steering robustness.
Therefore, the MDI-SM \red{provides a lower bound on} the degree of entanglement of the unknown quantum state and measurement incompatibility of the involved measurements. Besides, our approach is able to detect steerability in an MDI scenario with arbitrary detection losses and provide a lower bound on the steerability of the underlying assemblage.

Moreover, we tackle two optimization problems in Eq.~\eqref{Eq_SOMDI}. That is, the optimal measurement and MDI steering witness used for MDI-SM are obtained, or equivalently, we obtain the optimal strategies for the variant of QRSGs. At first glance, it seems to be a difficult problem to obtain the optimal measurement, since Bob has to optimize over all possible measurements. However, we show that the projection onto the maximally entangled state is always an optimal one for any steerable resource.
The optimal MDI steering witness (the variant QRSGs), on the other hand, can be efficiently computed by semidefinite programming. 
Finally, we provide an experimental demonstration of estimating the degree of steerability. \red{The result also bounds} the degree of entanglement, and incompatibility in an MDI scenario.
We have also proposed an improved MDI-SM which decreased the effect of some detection biases between Bob's detectors.

This work also reveals some open questions:
\red{It is interesting to investigate whether our method can be modified to all steerable assemblages in a standard DI scenario with the novel approach recently proposed in Refs.~\cite{Bowles18,Chen2020}. More recently, the DI certification of all steerable states has experimentally been implemented by self-testing an ancilla entangled pair~\cite{Zhao2019}.
It is also interesting to propose practical applications with the MDI scenario (or even a fully DI scheme following the work of Refs.~\cite{Bowles18,Chen2020,Zhao2019}). Since the formulation of the standard steering scenario can be applied to certify the security of quantum keys~\cite{Branciard12}, one can ask if this is also the case in the MDI scenario.}

\section*{Data and code availability}
The main data and code supporting the findings of this study are available within the manuscript. Additional data can be provided upon request from the corresponding author.

\begin{acknowledgments}
The authors acknowledge fruitful discussions with Francesco Buscemi, Ana Cristina Sprotte Costa, Yeong-Cherng Liang, Chau Nguyen, Paul Skrzypczyk, Roope Uola and Kang-Da Wu.

The authors acknowledge the support of the Graduate Student Study Abroad Program (Grant No. MOST 107-2917-I-006-002) for HYK; the Postdoctoral Research Abroad Program (Grant No. MOST 107-2917-I-564 -007) for SLC; the National Center for Theoretical Sciences and Ministry of Science and Technology, Taiwan (Grants Nos. MOST 107-2628-M-006-002-MY3, 108-2627-E-006-001, and 108-2811-M-006-536), and Army Research Office (Grant No. W911NF-19-1-0081) for YNC; the National Center for Theoretical Sciences and Ministry of Science and Technology, Taiwan (Grant No. MOST 108-2112-M-006-020-MY2) for HBC; the National Natural Science Foundation of China (Grants No. 11574291 and No. 11774334) for GYX; YYZ is supported by the National Natural Science Foundation for the Youth of China (No. 11804410); F.N. is supported in part by: NTT Research, Army Research Office (ARO) (Grant No. W911NF-18-1-0358), Japan Science and Technology Agency (JST) ( the CREST Grant No. JPMJCR1676), Japan Society for the Promotion of Science (JSPS) (via the KAKENHI Grant No. JP20H00134, and the grant JSPS-RFBR Grant No. JPJSBP120194828), and the Foundational Questions Institute (FQXi) (Grant No. FQXi-IAF19-06).
\end{acknowledgments}

\section{CONTRIBUTIONS:}

HYK and YYZ contributed equally to this work.
HYK and SLC contributed equally to the development of the theoretical analysis and conceived the project; GYX supervised the experiment; GYX and YYZ designed the experiment; YYZ conducted the experiment and collected data with the help from GYX; YYZ and HYK analyzed the experimental data with the help from GYX, CFL and GCG; HYK, SLC and HBC proved the theoretical results; YNC and FN supervised the research. All authors contributed to the writing of the manuscript.

\section{Competing interests:} The authors declare no competing interests.




\section{Figure legends}
Fig1: \textbf{Schematic illustration of the entanglement, quantum steering, Bell nonlocality, and MDI steering scenarios.} A pair of entangled photons $\rab$ (pink balls) are shared between two spatially separated parties: Alice and Bob. They verify whether they share the entanglement, steering, and nonlocal resource by violating the entanglement witness, steering inequality, and Bell inequality, respectively. (a) In the entanglement certification task, Alice and Bob both perform characterized measurements (transparent box). (b) In the quantum steering scenario, one party performs uncharacterized measurements (black box) according to the classical input $\{x\}$, while the other party performs a set of characterized measurements. (c) In Bell nonlocality, Alice (Bob) receives the classical input $\{x\}$ ($\{y\}$) and returns the outcomes $\{a\}$ ($\{b\}$) with uncharacterised measurements. (d) In the MDI steering scenario, Bob's classical input $\{y\}$ of the steering scenario is replaced with quantum inputs $\{\tau_y\}$, removing the necessity of trustiness of the measurement device.

Fig2: \textbf{Schematic drawing of the experimental setup.} (a) The singlet state of a pair of photons $\ket{HV}-\ket{VH}$ is generated by a spontaneous parametric down-conversion process, where $H$ ($V$) \red{represents} the horizontally (vertically) polarized direction. The Werner state is prepared by adding white noise (denoted by $\Omega$) to the system. Then one of the photons is sent to Alice, who uses Q1, H1, and PBS to perform the measurement $x$. The other photon is sent to Bob with an additional qubit system $\tau_y$ encoded on the photon's path degree of freedom `0' and `1'. We emphasize the preparation of the trusted quantum system in \green{panel (b)}.  Now Bob performs a complete Bell-state measurement on the equivalent two-qubit systems, i.e., measuring the polarization directions and the spatial paths of the \green{single particle}, and returns an outcome $b$. At the end, a set of probability distributions $\{p(a,b|x,\tau_y)\}$ is obtained to quantify the degree of steerability of the steerable resource. Abbreviations of the components are: BBO, barium borate crystal; HWP(H), \green{half-wave} plate; IF, interference filter; Att, attenuator; Mir, mirror; QP, quartz plate; QWP(Q), \green{quarter-wave} plate; PBS, polarizing beam splitter; BS, beam splitter; BD, beam displacer. The star represents that the HWP's axis is oriented at $45^{\circ}$.

Fig3: \textbf{Results of the MDI-SM and the estimation of entanglement and measurement incompatibility.} (a) The MDI experimental demonstration of estimating steerability of the family of two-qubit Werner states when considering Alice has three measurement settings. The theoretical prediction of the MDI-SM is plotted in the black line. The tailored estimator $\QMDI$ described in Eq.~\eqref{Eq_SOMDI_qubit} for this experiment is marked as diamonds ({\color{black} $\blacklozenge$}). The MDI-SM in Eq.~\eqref{Eq_SMDI} are marked using circles ({\color[rgb]{0, 1, 1} $\bigcirc$}), crosses ({\color[rgb]{0, 0, 1} $\times$}), stars ({\color[rgb]{0, 1, 0} $*$}), and triangles ({\color[rgb]{1,0,1} $\triangledown$}). (b) \red{MDI lower bounds on the degree of entanglement and incompatibility}. The diamond symbols ({\color{black} $\blacklozenge$}) in (a) and (b) represent the same quantity. We use the tailored estimator $\QMDI$ as lower bounds on the entanglement robustness (ER) of the underlying state and the incompatibility robustness (IR) of Alice's measurements. The actual values of these two quantities are represented by triangles ({\color{red} $\blacktriangledown$}) and squares ({\color{blue} $\blacksquare$}), respectively. By using the Monte Carlo algorithm, we obtain the standard deviations of $\SbMDILB$ in the value around $0.007$ and the standard deviations of $\QMDI$ in the value around  $0.004$ for three measurement settings by error propagation.

\end{document}